\documentclass[twoside]{dis07}
\usepackage[latin1]{inputenc}
\usepackage[dvips]{graphicx,epsfig,color}
\usepackage{wrapfig,rotating}
\usepackage{amssymb,amsmath,array}

\begin{document}
\title{Inclusive Jet Production in DIS at High $Q^{2}$ and Extraction of the Strong Coupling}

\author{Thomas Kluge
\thanks{on behalf of the H1 Collaboration}
%
\vspace{.3cm}\\
%
DESY \\
Notkestr. 85, 22607 Hamburg - Germany
%
}

\maketitle

\begin{abstract}
Inclusive jet production is studied in neutral
current deep-inelastic positron-proton scattering at large
four momentum transfer squared $Q^2>150{\,\mbox{GeV}}^2$ with the H1
detector at HERA.
The measurements are found to be well described by
calculations at next-to-leading order in perturbative QCD.
 The running of the strong coupling is demonstrated  
and the value of $\alpha_s(M_Z)$ is determined.
\end{abstract}

\section{Introduction and Experimental Method}
Jet production in neutral current (NC) deep-inelastic
scattering (DIS) at HERA provides an important testing ground for Quantum Chromodynamics (QCD).
The Born contribution in DIS
gives only indirect information on the strong coupling $\alpha_s$ 
via scaling violations of the proton structure functions.
At leading order (LO) in $\alpha_s$ additional processes contribute:
QCD-Compton and boson-gluon fusion.

In the Breit frame of reference~\cite{feynman},
where the virtual boson and the proton collide head on,
 the Born contribution generates no transverse momenta.
Partons with transverse momenta are produced in lowest order by
 the QCD-Compton and boson-gluon fusion processes.
Jet production in the Breit frame therefore provides direct sensitivity to
 $\alpha_s$ and allows for a precision test of QCD.

In this workshop contribution new measurements of the inclusive jet cross section are presented, 
 based on data corresponding to  twice the integrated luminosity and
a higher centre-of-mass energy than in the previous H1 analysis~\cite{Adloff:2000tq}.
The larger data set together with improved understanding of the hadronic energy measurement significantly reduces the total 
uncertainty of the results.
The data were collected with the H1 detector at HERA in the years 1999 and 2000.
During this period HERA collided positrons of energy $E_e=27.5{\,\mbox{GeV}}$ with protons of energy $E_p=920{\,\mbox{GeV}}$
giving a centre-of-mass energy $\sqrt{s} = 319{\,\mbox{GeV}}$.
The data sample used in this analysis corresponds to an integrated luminosity of $65.4{\,\rm pb}^{-1}$.

The DIS phase space covered by this analysis is defined by $150<Q^2<15000~{\,\mbox{GeV}}^2 \ ,$ $0.2<y<0.7 \ ,$
where $y$ quantifies the inelasticity of the interaction.
These two variables are reconstructed from the four momenta of the scattered positron and the
hadronic final state particles using the electron-sigma method~\cite{Bassler:1994uq}.

The jet analysis is performed in the Breit frame.
The boost from the laboratory system to the Breit frame is determined by $Q^2$, $y$ and
 the azimuthal angle of the scattered positron.
Particles of the hadronic final state are clustered
into jets using the inclusive $k_T$ algorithm~\cite{Ellis:1993tq}
with the $p_T$ recombination scheme and with distance parameter $R=1$ in the \mbox{$\eta$-$\phi$} plane.
The inclusive $k_T$ algorithm is infrared safe and results in small hadronisation corrections \cite{Adloff:2000tq}. 
Every jet with $7 < E_T < 50{\,\mbox{GeV}}$ contributes to the inclusive jet cross section,
 regardless of the jet multiplicity in the event.
In addition, the normalised inclusive jet cross section is investigated,
calculated as the ratio of the number of jets to the number of selected NC DIS events
in the $y$~range defined above.
This observable equals the average jet multiplicity of NC DIS events within the given phase space.
Jet cross sections and normalised jet cross sections are studied as a function of $Q^2$ and $E_T$.

The following sources of systematic uncertainty are considered:
 positron energy uncertainty  ($0.7\%$~to~$3\%$ depending on the $z$-impact point of the positron in the calorimeter),
 positron polar angle systematic uncertainty ($1$~and~$3~\mathrm{mrad}$),
 energy scale uncertainty of the reconstructed hadronic final state ($2\%$),
 luminosity measurement uncertainty ($1.5\%$).
The model dependence of the data correction is below $10\%$ in most of the bins and typically $2\%$.
An error of $1\%$ is estimated from QED radiative correction uncertainty.
The dominant experimental uncertainties on the jet cross section arise from the model dependence of the data correction
 and from the LAr hadronic energy scale uncertainty.
 The individual contributions are added in 
quadrature to obtain the total systematic uncertainty.
The correlations of the errors among the different bins are taken into account.
For the normalised jet cross sections
systematic uncertainties are reduced and the luminosity uncertainty cancels.

The theoretical prediction for the jet cross section is obtained using the  NLOJET++ program~\cite{Nagy:2001xb}, which
performs the matrix element integration at NLO of the strong coupling, $\mathcal{O}(\alpha_s^2)$.
The strong coupling is taken as $\alpha_s(M_Z) = 0.118$  and is evolved as a function of the renormalisation scale at two loop precision.
The calculations are performed in the $\overline{\mbox{\rm MS}}$ scheme for five massless quark flavours.
The parton density functions (PDFs) of the proton are taken from the CTEQ6.5M set~\cite{Tung:2006tb}.
The factorisation scale $\mu_f$ is chosen to be $Q$  and the renormalisation scale $\mu_r$ is chosen to be the $E_T$ of each jet. 
Running of the electromagnetic coupling with $Q^2$ is taken into account.
In order to calculate the normalised inclusive jet cross sections,
the prediction of the inclusive jet cross section is divided by the prediction of the NC DIS cross section.
The latter is calculated at NLO, $\mathcal{O}(\alpha_s)$, with the DISENT package \cite{Catani:1996vz},
and the renormalisation and factorisation scales set to $Q$.
The strong coupling is determined by repeating the perturbative calculations for many values of $\alpha_s(M_Z)$
 until the best match of data and theory is found.
With NLOJET++ and DISENT these calculations are prohibitively time consuming.
A considerable gain in computational speed is provided by the fastNLO package~\cite{Kluge:2006xs}.
All theory calculations shown in the following are obtained using fastNLO.

\section{Results}

\begin{figure}[h]
\begin{center}
\includegraphics[width=0.50\columnwidth]{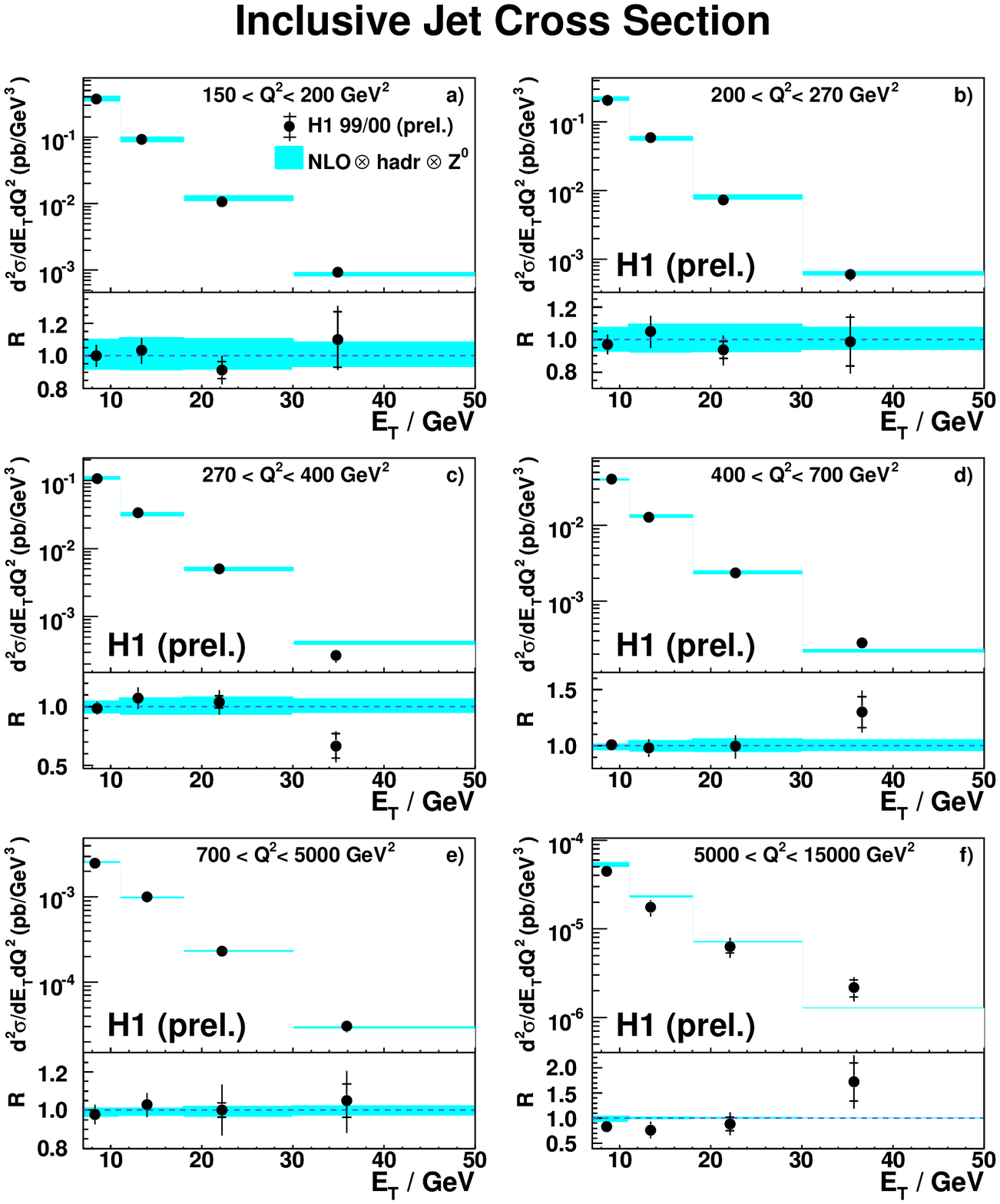}
\includegraphics[width=0.48\columnwidth]{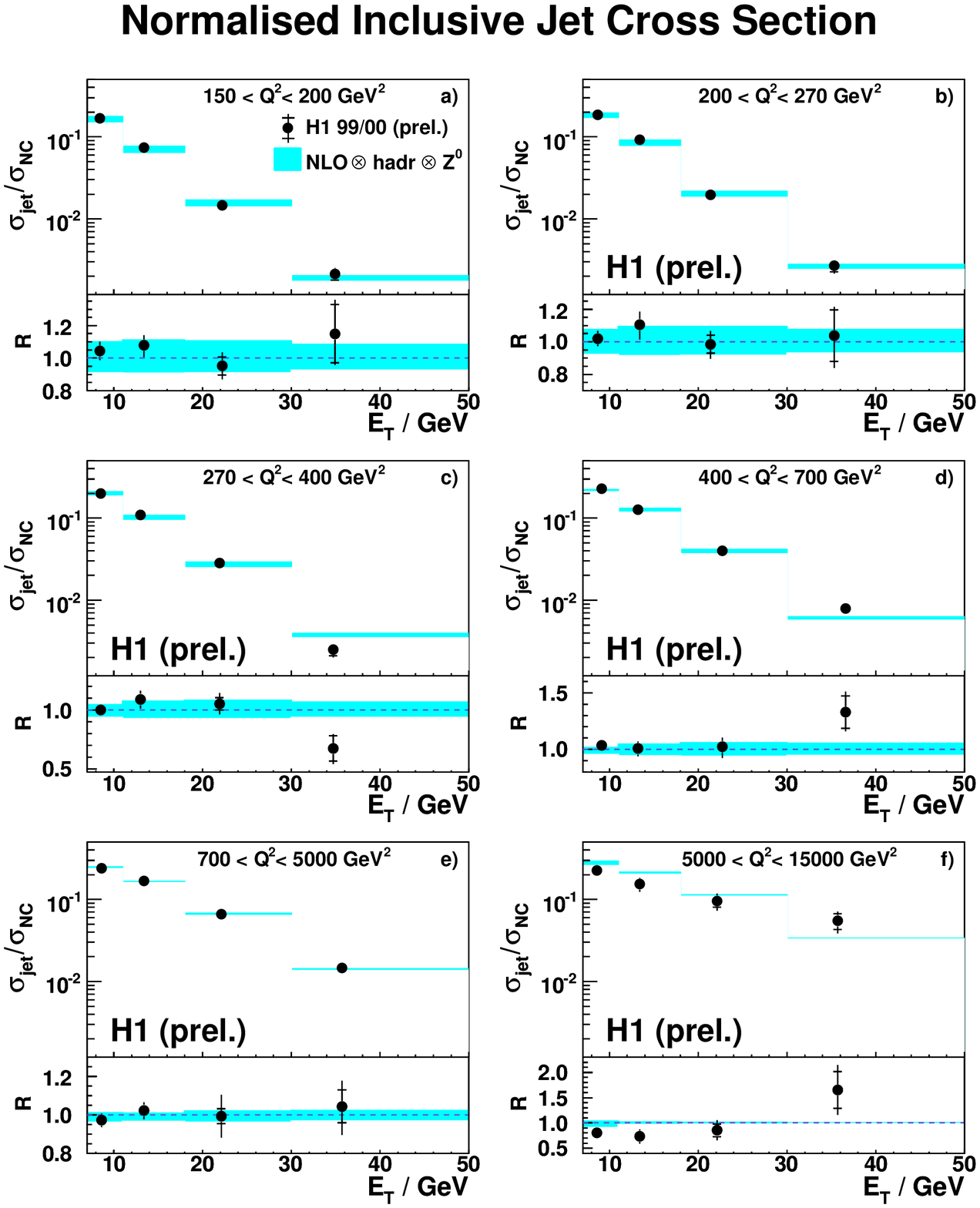}
\end{center}
\caption{The double differential cross section as a function of $E_T$ for six regions of $Q^2$.
The data, presented with statistical errors (inner bars)
 and total errors (outer bars), are compared with the results of NLOJET++, corrected
for hadronisation and $Z^0$ boson exchange. 
}\label{fig:sigma}
\end{figure}

The measured cross sections, corrected for detector
and radiative QED effects, are presented as double differential distributions in figure \ref{fig:sigma}.
The data points are shown at the average value of the $Q^2$ or $E_T$ in each bin.
The results are compared to the perturbative QCD predictions in NLO with $\alpha_s(M_Z) = 0.118$,
 taking into account hadronisation effects and $Z^0$-exchange.
The inclusive jet cross section is shown in figure~\ref{fig:sigma} (left) as a function of $E_T$ in
six $Q^2$ bins in the range $150 < Q^2 < 15000\,\mathrm{ GeV^2} $.
The data are well described by the theory over the full $E_T$ and $Q^2$ ranges, with $\chi^2/\mathrm{ndf} = 16.7/24$,
 taking only experimental errors into account.

For NC DIS events in the range $0.2<y<0.7$ and in a given $Q^2$ bin
the normalised inclusive jet cross section is defined as the average number of jets within $-1.0 < \eta^\mathrm{Lab} < 2.5$
per event.
Figure~\ref{fig:sigma} (right) shows the normalised inclusive jet cross section as a function of $E_T$ in six $Q^2$ bins.
The NLO calculation gives a good description of the data in the full $E_T$ and $Q^2$ range.   
Compared with the inclusive jet cross section, the normalised 
inclusive jet cross section exhibits a smaller experimental uncertainty.

\begin{figure}
\begin{center}
\includegraphics[width=0.7\columnwidth]{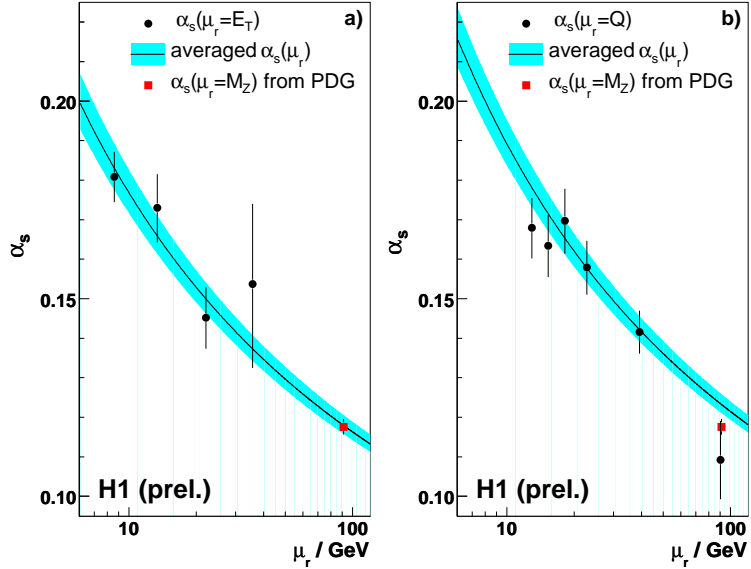}
\end{center}
\caption{Results for the fitted values of a) $\alpha_s(\mu_r=E_T)$ averaged over all $Q^2$ regions, and b) 
 $\alpha_s(\mu_r=Q)$ averaged over all $E_T$ regions.
The error bars denote the total experimental uncertainty for each data point.
The solid curve shows the result of evolving $\alpha_s(M_Z)$ averaged from all $Q^2$ and $E_T$ regions, 
 with the band denoting the total experimental uncertainty.
}\label{fig:theoet}
\end{figure}

A fit of $\alpha_s(M_Z)$ to all of the 24 measurements of the double differential inclusive jet cross sections is made,
which yields $\alpha_s(M_Z) = 0.1179 ~\pm 0.0024\,\mathrm{(exp.)}
~ ^{+0.0052}_{-0.0032}\,\mathrm{(th.)}~ \pm 0.0028\,\mathrm{(pdf)} ,
\label{alphas1} $ with a fit quality: $\chi^2/\mathrm{ndf} = 20.2/23$.
To study the scale dependence of $\alpha_s$, the six data points at a given
$E_T$ are used together, and four values of $\alpha_s(E_T)$ are extracted.
The results are shown in figure~\ref{fig:theoet}a,
 where the running of the strong coupling  is also clearly observed.
In figure~\ref{fig:theoet}b the results using an alternative scale $Q$ instead of $E_T$ are shown,
the four data points at a given
$Q^2$ are used together, and six values of $\alpha_s(Q)$ are extracted.
These results are larger but compatible with the values obtained at the scale $E_T$.

The strong coupling is also fitted to the normalised inclusive jet cross section.
All 24 measurements are used in a common fit, 
which yields
$$\alpha_s(M_Z) = 0.1193 ~\pm 0.0014\,\mathrm{(exp.)}~ ^{+0.0047}_{-0.0030}\,\mathrm{(th.)}~ \pm 0.0016\,\mathrm{(pdf)} ,$$
with a fit quality of $\chi^2/\mathrm{ndf} = 28.7/23$.
This result is compatible within errors with the value from the inclusive jet cross sections.
The normalisation gives rise to cancellations of systematic effects, which lead to improved experimental and PDF uncertainties.
This determination of $\alpha_s(M_Z)$ is consistent with the world 
average $\alpha_s(M_Z)=0.1176 \pm 0.0020$ ~\cite{Yao:2006px} and with the previous H1
determination from inclusive jet production measurements~\cite{Adloff:2000tq}.
The dominating theory error can be reduced at the expense of a larger experimental uncertainty
by restricting the phase space of the fit interval to higher values of $Q^2$.
The smallest total uncertainty is obtained by a combined fit
 of the normalised inclusive jet cross section for $700<Q^2<5000\,\mathrm{ GeV^2}$:
$ \alpha_s(M_Z) = 0.1171 ~\pm0.0023\,\mathrm{(exp.)}
~ ^{+0.0032}_{-0.0010}\,\mathrm{(th.)}~ \pm0.0010\,\mathrm{(pdf)}$,
with a fit quality of $\chi^2/\mathrm{ndf} = 1.2/3$.
This result shows a level of experimental precision competitive with $\alpha_s$ determinations from 
other recent jet production measurements at HERA~\cite{Chekanov:2006yc} and those from $e^+ e^-$ data~\cite{Abbiendi:2005vd}
and is in good agreement with the world average.

\begin{footnotesize}



\end{footnotesize}



\begin{thebibliography}{99}
\bibitem{url} Slides: \\ 
\verb$http://indico.cern.ch/contributionDisplay.py?contribId=214&sessionId=6&confId=9499$
\bibitem{feynman}
  R.~P.~Feynman,
``Photon-Hadron Interactions'',
  Benjamin, New York (1972).
\bibitem{Adloff:2000tq}
  C.~Adloff {\it et al.},
  Eur.\ Phys.\ J.\ C {\bf 19} 289 (2001).
\bibitem{Bassler:1994uq}
  U.~Bassler and G.~Bernardi,
  Nucl.\ Instrum.\ Meth.\ A {\bf 361} 197 (1995).
\bibitem{Ellis:1993tq}
  S.~D.~Ellis and D.~E.~Soper,
  Phys.\ Rev.\ D {\bf 48} 3160 (1993).
\bibitem{Nagy:2001xb}
  Z.~Nagy and Z.~Trocsanyi,
  Phys.\ Rev.\ Lett.\  {\bf 87} 082001 (2001).
\bibitem{Tung:2006tb}
  W.~K.~Tung {\it et al.},
  JHEP {\bf 0702} 053 (2007).
\bibitem{Catani:1996vz}
  S.~Catani and M.~H.~Seymour,
  Nucl.\ Phys.\  B {\bf 485} 291 (1997)
  [Erratum-ibid.\  B {\bf 510} 504 (1998)].
\bibitem{Kluge:2006xs}
  T.~Kluge, K.~Rabbertz and M.~Wobisch,
  arXiv:hep-ph/0609285 (2006).
\bibitem{Yao:2006px}
  W.~M.~Yao {\it et al.},
  J.\ Phys.\ G {\bf 33} 1 (2006).
\bibitem{Chekanov:2006yc}
  S.~Chekanov {\it et al.},
  Submitted to Phys.\ Lett.\ B,
  arXiv:hep-ex/0701039 (2007).
\bibitem{Abbiendi:2005vd}
  G.~Abbiendi {\it et al.},
  Eur.\ Phys.\ J.\  C {\bf 47} 295 (2006).


\end{thebibliography}
\end{document}